\documentclass{aastex}          
\usepackage{spr-astr-addons}    

\begin{document}
%
\title{IMF biases created by binning and unresolved systems}

\shorttitle{IMF biases created by binning and unresolved systems}
\shortauthors{Ma\'{\i}z Apell\'aniz}

\author{J. Ma\'{\i}z Apell\'aniz\altaffilmark{1,2,3}} 

\altaffiltext{1}{IAA-CSIC, Granada, Spain}
\altaffiltext{2}{Ram\'on y Cajal Fellow}
\altaffiltext{2}{Support for this work was provided by the Spanish Government Ministerio de Educaci\'on y Ciencia through grants AYA2004-08260-C03 and AYA2007-64712.}

\begin{abstract}
I discuss two of the possible sources of biases in the determination of the IMF: binning and the existence of unresolved components. 
The first source is important for clusters with a small number of stars detected in a given mass bin while the second one is relevant for 
all clusters located beyond the immediate solar neighborhood. For both cases I will present results of numerical simulations 
and I will discuss strategies to correct for their effects. I also present a brief description of a third unrelated bias source.
\end{abstract}


%
\section{Introduction}

	Most initial mass functions (IMFs) of simple stellar populations are calculated by
positioning the observational data in a color-magnitude (CMD) or in a Hertzsprung-Russell (HRD) diagram. From stellar models one 
extracts the isochrone appropriate for the age and metallicity and compares it with the data to obtain the evolutionary masses for the 
individual stars. Stars are then grouped into bins with equal width in $\Delta(\log m)$ and a function (e.g. 
$dn/dm = Am^\gamma$) is fitted to the binned data using $\chi^2$ minimization.

	The above procedure is subject to different biases that can influence the derived parameters. In this
contribution, I discuss two of them and present my plans to analyze a third. The first bias is the one induced by using bins of equal 
width when fitting power-laws to binned data. Such an effect is more general than its application to the calculation of mass functions
and was discussed in Paper I \citep{MaizUbed05}. The second bias is caused by the existence of unresolved multiple systems, either physical
or chance alignments, and was discussed in Paper II \citep{Maiz08a}. The third bias is related to the effect of random uncertainties in 
the mass determinations on the computed IMF and will be analyzed in a future paper.

\section{Binning biases}

	It is well known that when data are binned and a function is fitted to the outcome a bias in the derived parameters can be (and 
usually is) present if there is a low number of objects in some of the bins (see, e.g. \citealt{BeviRobi92,NousShue89}). The problem 
originates in the strong anticorrelation between the data and the weights in chi-square minimization for data with Poisson or
binomial uncertainties \citep{Wheaetal95}. In the case of IMF calculations using equal-width bins biases can be large because the
values of the slope $\gamma$ are typically large and at the high end there are usually only a handful of stars.

%
\begin{figure}[t]
\includegraphics[width=\columnwidth]{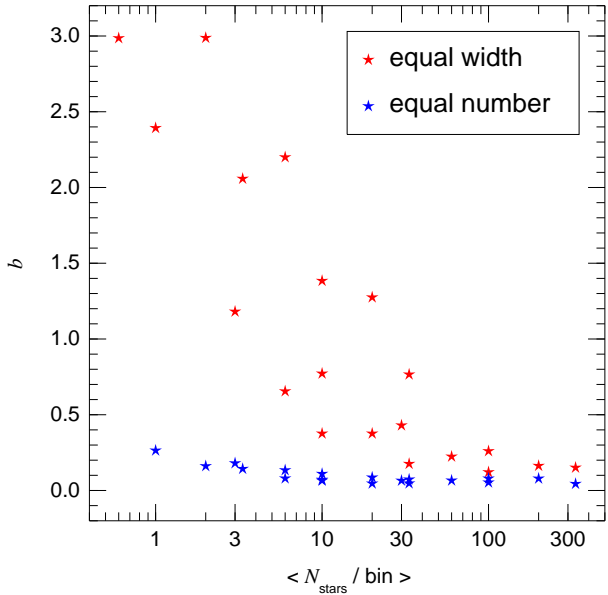}
\caption{Bias as a function of the mean number of stars per bin for the equal-width and the equal-number experiments.} 
\label{biasplot}
\end{figure}

	There are several ways to deal with this bias. In Paper I we introduced the idea of binning the data by placing the same number of
stars in each bin. Such a technique shifts the information on the IMF from the number of stars per bin to the positions of the limits
between bins. Since each bin has then the same weight\footnote{If the total number of stars divided by the number of bins is not an 
integer, one has to place one more star in some bins compared to others, leading to slightly different weights. Nevertheless,
such an effect is very small in most cases.}, there is no anticorrelation between the data and the weights, thus minimizing the biases.

	The idea of using equal-number bins instead of equal-width bins was tested in Paper I using Montecarlo simulations with different
total number of stars (30, 100, 300, or 1000) and number of bins (3, 5, 10, 30, 50). For each combination of numbers of stars and bins 
1000 realizations of a Salpeter IMF with $M > 6.3 $ M$_\odot$ were obtained and the slope of each one of them was measured with a 
$\chi^2$-fitting algorithm. Then, for each of the combinations the bias $b$ was defined as:

\begin{equation}
b = <(\gamma_k+2.35)/\sigma_k>,
\end{equation}

\noindent where $k$ is the realization running index and $\gamma_k$, $\sigma_k$ are the measured IMF slope and uncertainty, respectively,
for the $k$th realization.

	Results are shown in Figure~\ref{biasplot}. Fits with equal-number bins consistently yield smaller biases than fits with 
equal-width bins. The effect is especially large when the mean number of stars per bin is low: for values as low as 20, the bias can be
larger than 1 when equal-width bins are used. On the other hand, even when only one star per bin is used, the bias for the equal-number
case remains smaller than 0.3. Furthermore, we tested that the uncertainties produced by the $\chi^2$-fitting algorithm indeed correspond 
to the real uncertainties because the standard deviation of the distribution of $(\gamma_k+2.35)/\sigma_k$ is very close to 1.0 in all cases. 

	An interesting side outcome of our Montecarlo experiments is the determination of the minimum number of stars required to measure
the slope of an IMF given its expected random uncertainty. The lower the number of stars, the larger the uncertainty because sampling
effects become more important. We found the number to be $\sim$100 stars for $\sigma_k = 0.2$ and $\sim$30 stars for $\sigma_k = 0.3$. 

\section{Unresolved systems}

%
\begin{figure}[t]
\includegraphics[width=\columnwidth]{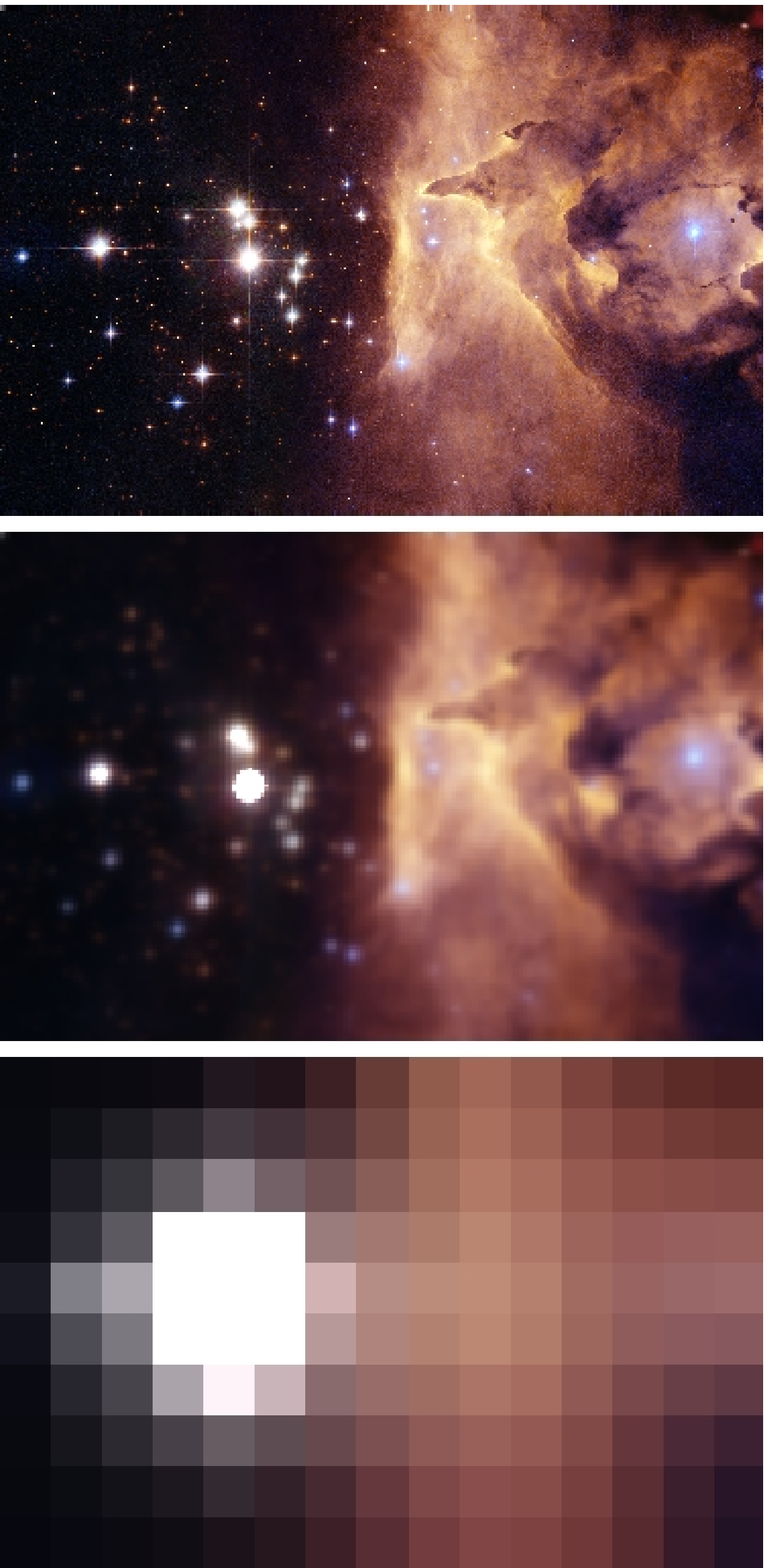}
\caption{[top] HST/ACS+WFPC2 3-color mosaic of Pismis~24, a stellar young cluster located at a distance of 2.5 kpc. The field size is
2.9 pc $\times$ 1.9 pc. [middle] A simulation of the same field observed with HST at the distance of the LMC. [bottom] A simulation of the 
same field observed with HST at the distance of M33. Note that the brightest apparent point source in the top frame is actually composed of 
three stars with masses above 50 M$_\odot$, of which two of them remain unresolved with HST at 2.5 kpc.}
\label{Pismis24}
\end{figure}

	It is believed that most (if not all) massive stars are formed in multiple systems and that only runaway objects have a relatively
low probability of having no companions \citep{Masoetal98}. Given the typical distance to massive young clusters (MYCs), most of those 
binaries will be unresolved even with HST. Also, given the distances and core densities of MYCs, chance superpositions between cluster 
members should be a common occurrence. An example of the importance of the problem is shown in Fig.~\ref{Pismis24}. 

	In Paper II I modeled the effect of real multiple systems and chance alignments on the observed color-magnitude diagrams. I also
studied how the apparent mass function (AMF) differs from the IMF if those effects are not taken into account. Below I summarize the main
results.

\subsection{Real multiple systems}

%
\begin{figure}[t]
\includegraphics*[width=\columnwidth, bb=28 28 566 530]{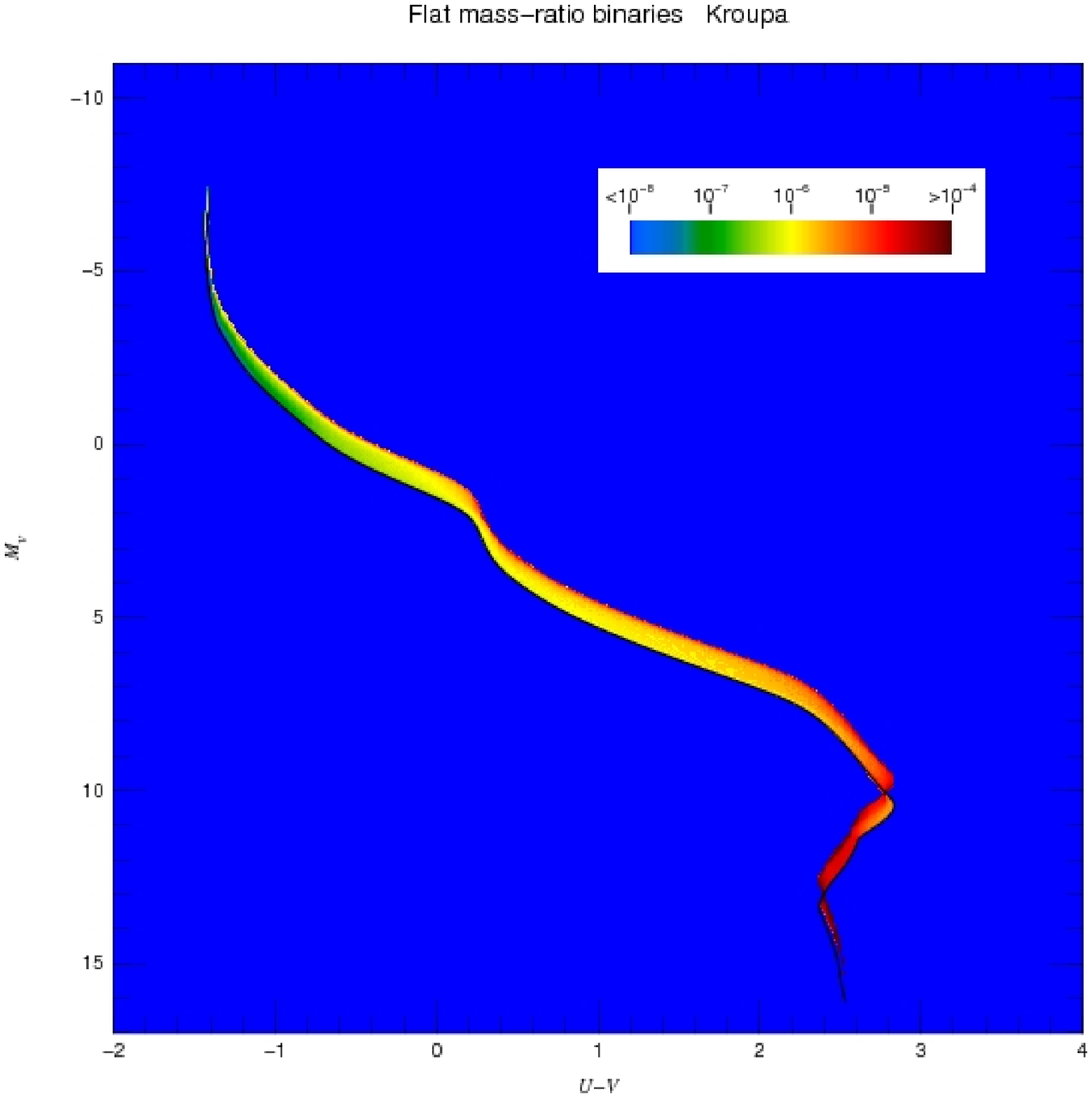}
\caption{[top] Color-magnitude density function for real binary systems with a Kroupa IMF and a flat mass ratio shown as a Hess diagram. 
The black line shows the position of the 1 million year isochrone. The function scaling is logarithmic and the normalization is arbitrary.}
\label{uvv_bin}
\end{figure}

%
\begin{figure}[t]
\includegraphics*[width=\linewidth, bb=28 28 566 530]{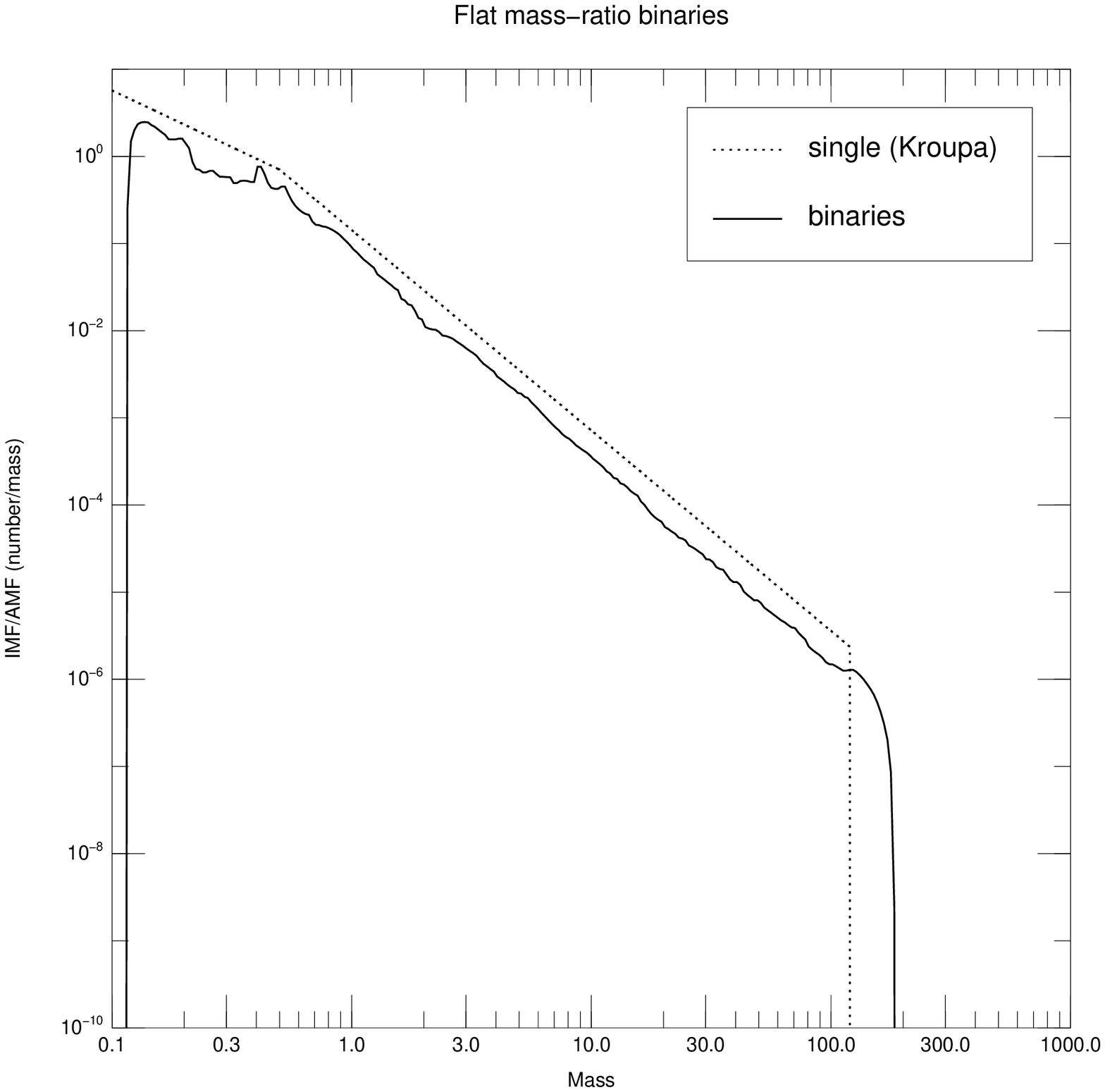}
\caption[top]{IMF and AMF for real binary systems with a Kroupa IMF and a flat mass ratio. The mass is
expressed in solar units. The IMF is normalized to 1.0 and the AMF to 0.5.}
\label{imf_bin}
\end{figure}	

	In the first experiment of Paper II, I modeled the case in which all stars in a cluster are unresolved binaries, the IMF for
individual stars is Kroupa or Kroupa-like, and the secondary mass function is flat (hence, for a given primary mass, the companion
will be, on average, more massive than if drawn from an IMF clipped at the primary mass). 
The observed CMD for a Kroupa IMF and a comparison between the IMF and
the AMF are shown in Figs.~\ref{uvv_bin}~and~\ref{imf_bin}. The observed CMD is widened due to the existence of binaries and a population
of apparently ultramassive stars is formed: if the real stellar upper mass limit is 120 M$_\odot$, the apparent one is 182 M$_\odot$. 
The effect on the mass function slope for massive stars is small, though, as the AMF and the IMF are nearly parallel in Fig.~\ref{imf_bin}.

\subsection{Chance alignments of single stars and binaries}

%
\begin{figure*}[b]
\centerline{\includegraphics*[angle=90, width=0.85\linewidth]{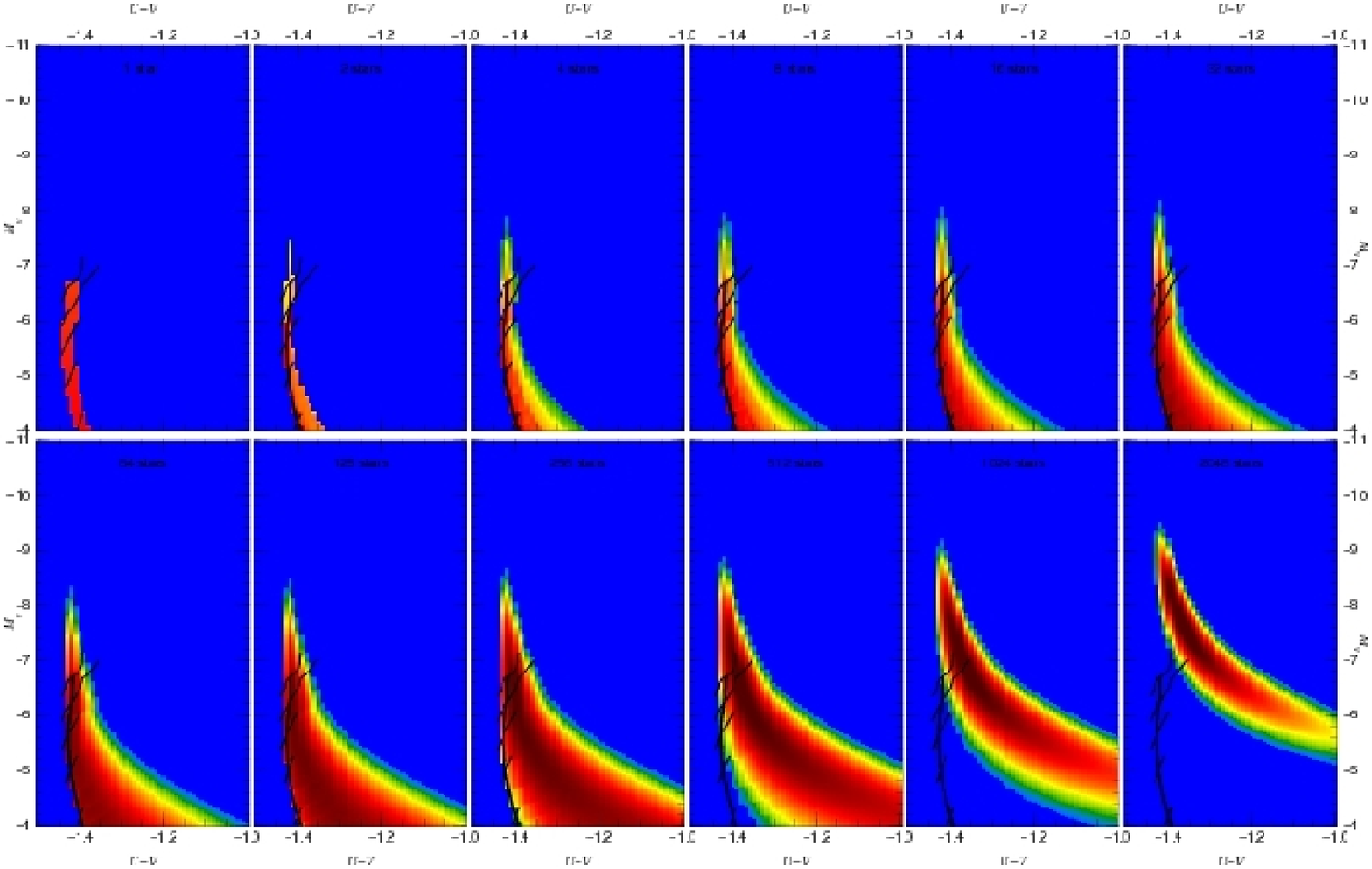}}
\caption[top]{Top left corner of the color-magnitude density functions for $N=1,2,4\ldots 2048$ chance superpositions of single stars and a
Kroupa IMF shown as Hess diagrams. The thick black line shows the position of the 1 Ma isochrone and the thin lines the evolutionary tracks
between 0 and 2 Ma for the initial masses of 25, 40, 60, 85, and 120 M$_\odot$. The function scaling is logarithmic and the normalization 
is arbitrary.}
\label{uvv_chsup}
\end{figure*}	

%
\begin{figure*}[t]
\centerline{\includegraphics*[angle=90, width=0.85\linewidth]{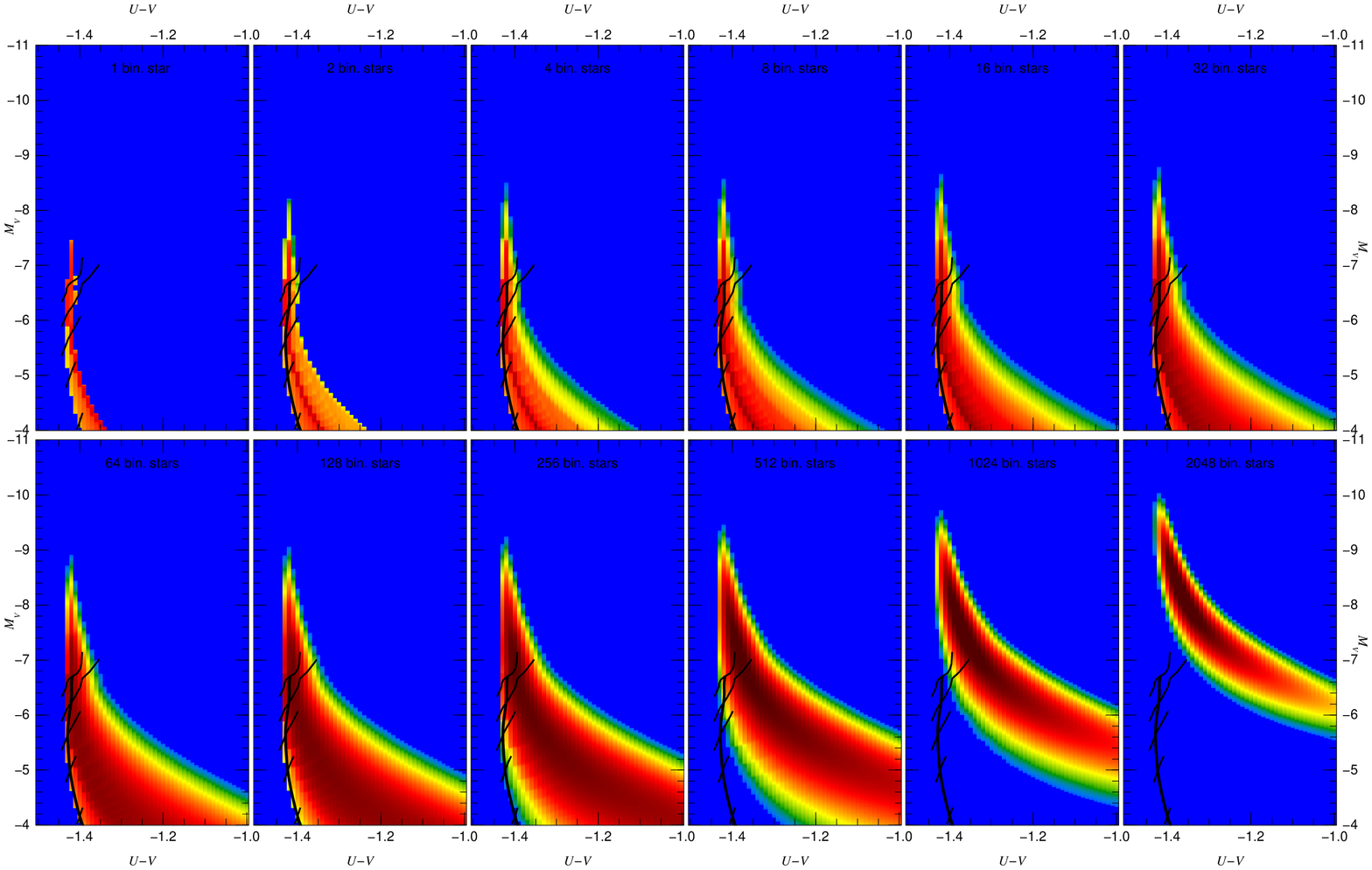}}
\caption[top]{Same as Fig.~\ref{uvv_chsup} for chance superpositions of binaries.}
\label{uvv_chsupbin}
\end{figure*}	

%
\begin{figure}[t]
\includegraphics*[width=\linewidth, bb=28 28 566 530]{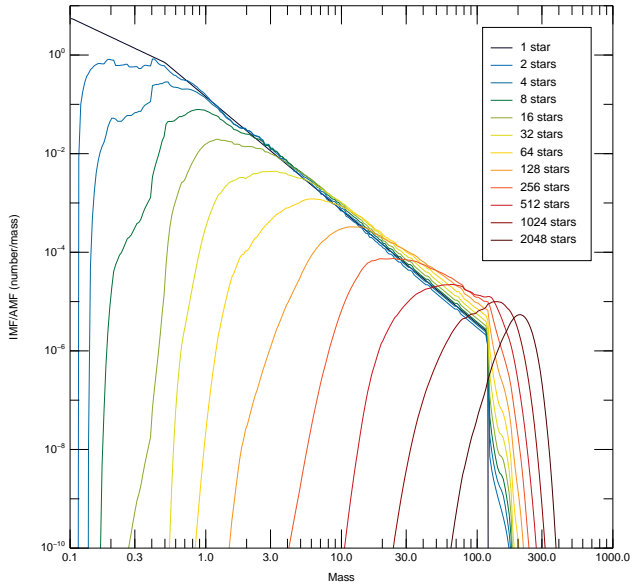}
\caption[top]{IMF and AMFs for chance superpositions of single stars with a Kroupa IMF. The mass is expressed in solar units.}
\label{imf_chsup}
\end{figure}	

%
\begin{figure}[t]
\includegraphics*[width=\linewidth, bb=28 28 566 530]{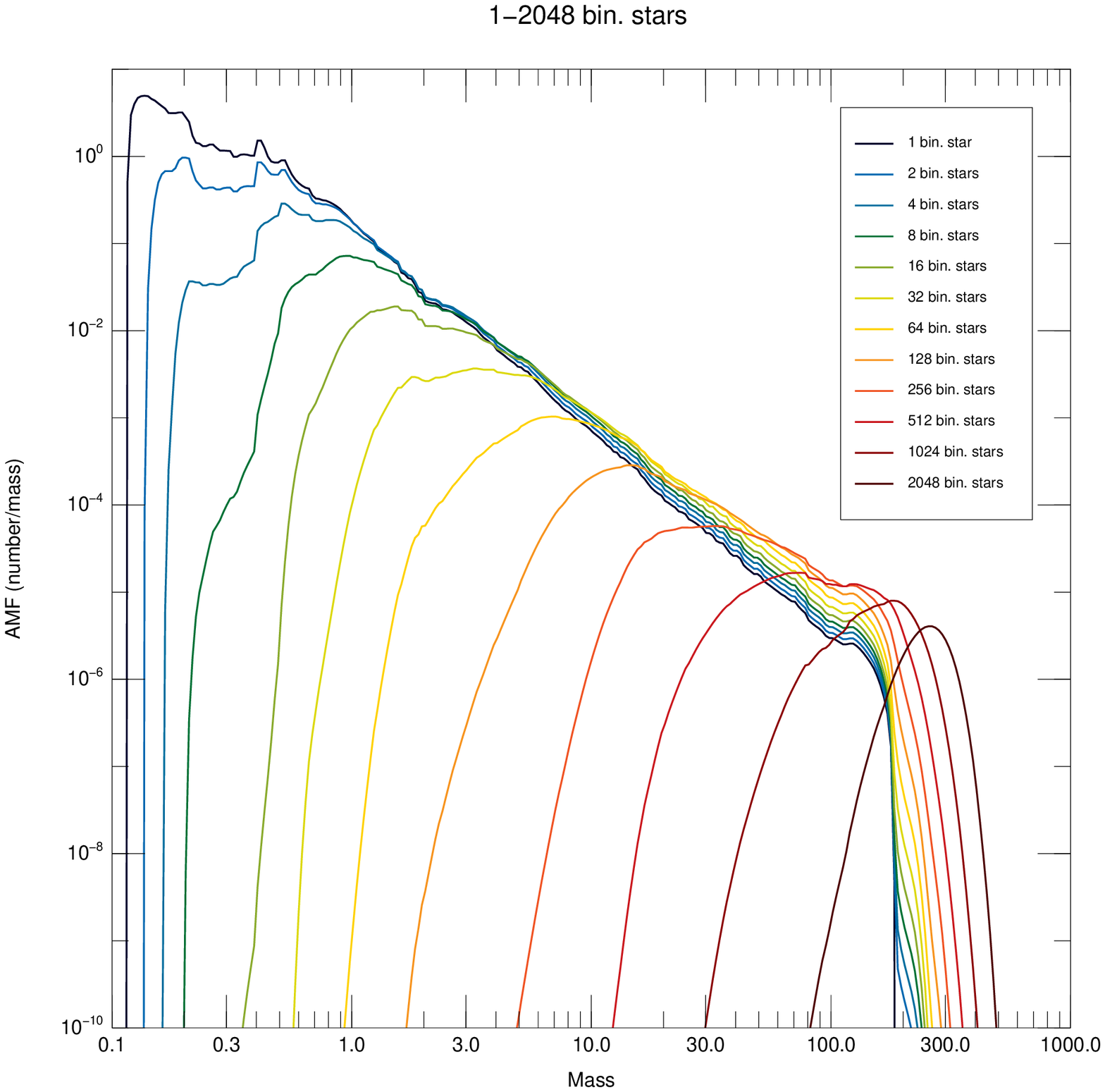}
\caption[top]{Same as Fig.~\ref{imf_chsup} for chance superpositions of binaries.}
\label{imf_chsupbin}
\end{figure}	

	In the second and third experiments of Paper II, I treated the effect of chance superpositions due to the increased crowding that 
takes place as the distance or central density of a cluster becomes larger. This was done by means of a ``fat pixel'' approximation that 
considers that the light from a given star falls on a single detector pixel. As crowding increases, each pixel includes light from more 
stars and the observed CMD departs more from the original one. The second experiment analyzed the cases where $N=1, 2, 4\ldots 2048$ single
stars are included in each detected source. The third experiment did the same using binaries instead of single stars. The blue and 
luminous part of the observed CMD are shown in Figs.~\ref{uvv_chsup}~and~\ref{uvv_chsupbin} and the derived AMFs are shown in 
Figs.~\ref{imf_chsup}~and~\ref{imf_chsupbin}. 
	
	The observed CMD for the second experiment show that, as the number of superimposed stars increases, the area populated with 
objects becomes first wider and then narrows down. As it happened with the first experiment, a population of apparently ultramassive stars 
also forms but it takes a relatively large number of superimposed stars for its number to become significant. At the other end, the
increase in the number of superpositions implies that we lose our ability to detect low-mass stars. The results for the third experiment
are similar except that for the case where $N=1$ (which is equivalent to the first experiment) a significant number of apparently
ultramassive stars already exists.

	With respect to the slope of the AMF for massive stars for the second and third experiments, there are three regimes. For $N$
small, the change in the slope is small ($<0.2$), has a weak dependence on $N$, and is easily correctable. For intermediate values of $N$
($\sim$ 30), the change in the slope is larger and more dependent on $N$ but a correction can still be used. For large values of $N$ the
situation is hopeless and the true IMF cannot be measured from the data.

\subsection{Sample applications: NGC 3603 and R136}

%
\begin{figure}[t]
\includegraphics*[width=\linewidth, bb=24 24 570 498]{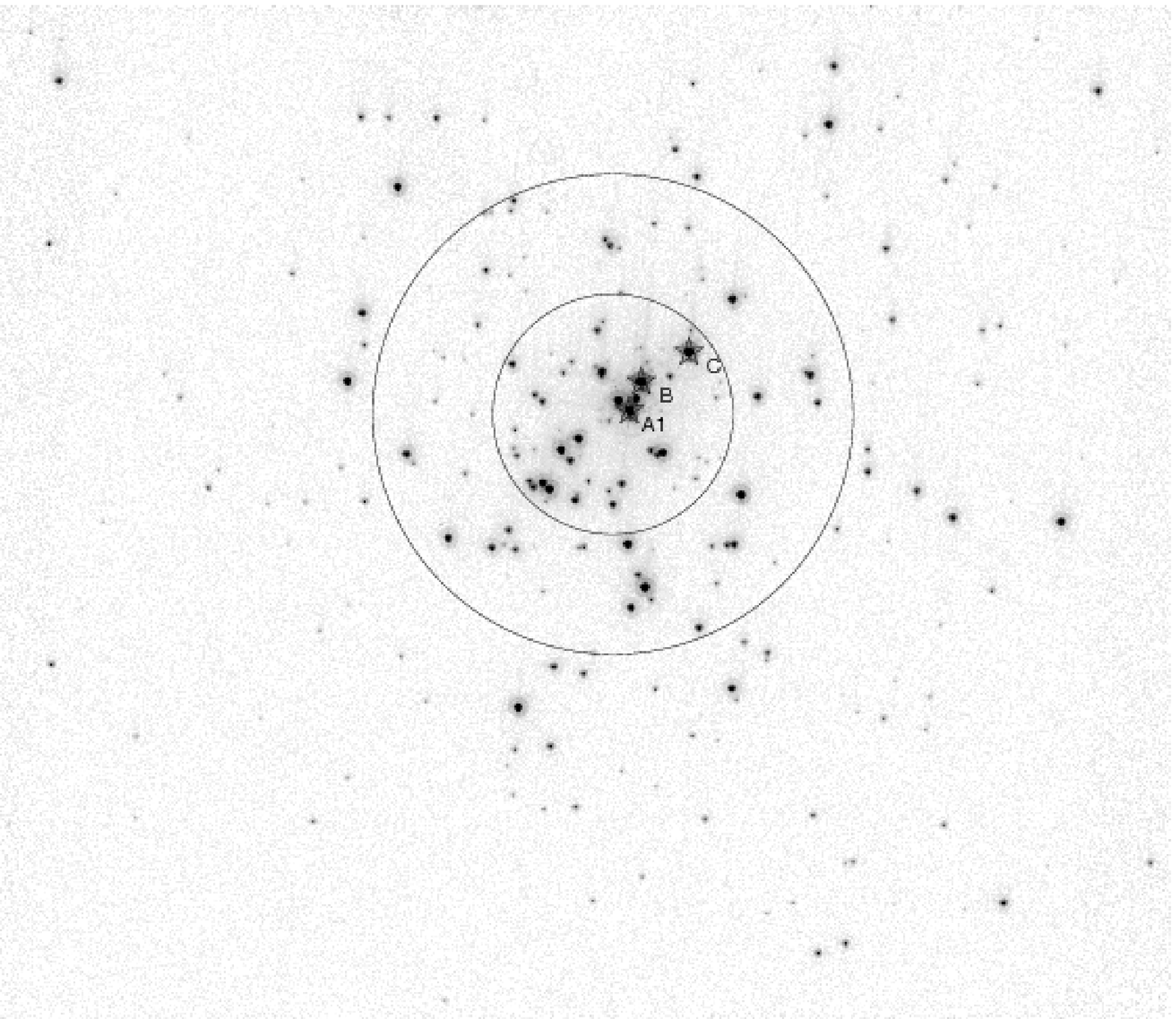}
\caption[top]{ACS/HRC F550M DRZ frame of NGC 3603. The three most massive (WNha) objects are marked and circles with radii of 3\arcsec\ and 
6\arcsec\ centered on the cluster have been drawn. A square-root scale between 0 and 800 counts has been used in order to show both the bright and 
the dim stars. The pixel size is 0\farcs025, the field size is 29\farcs3$\times$25\farcs4 (1172 px $\times$ 1016 px), and the vertical 
direction is 124\degr\ East of North.}
\label{ngc3603_f550m}
\end{figure}	

%
\begin{figure}[t]
\includegraphics*[width=\linewidth]{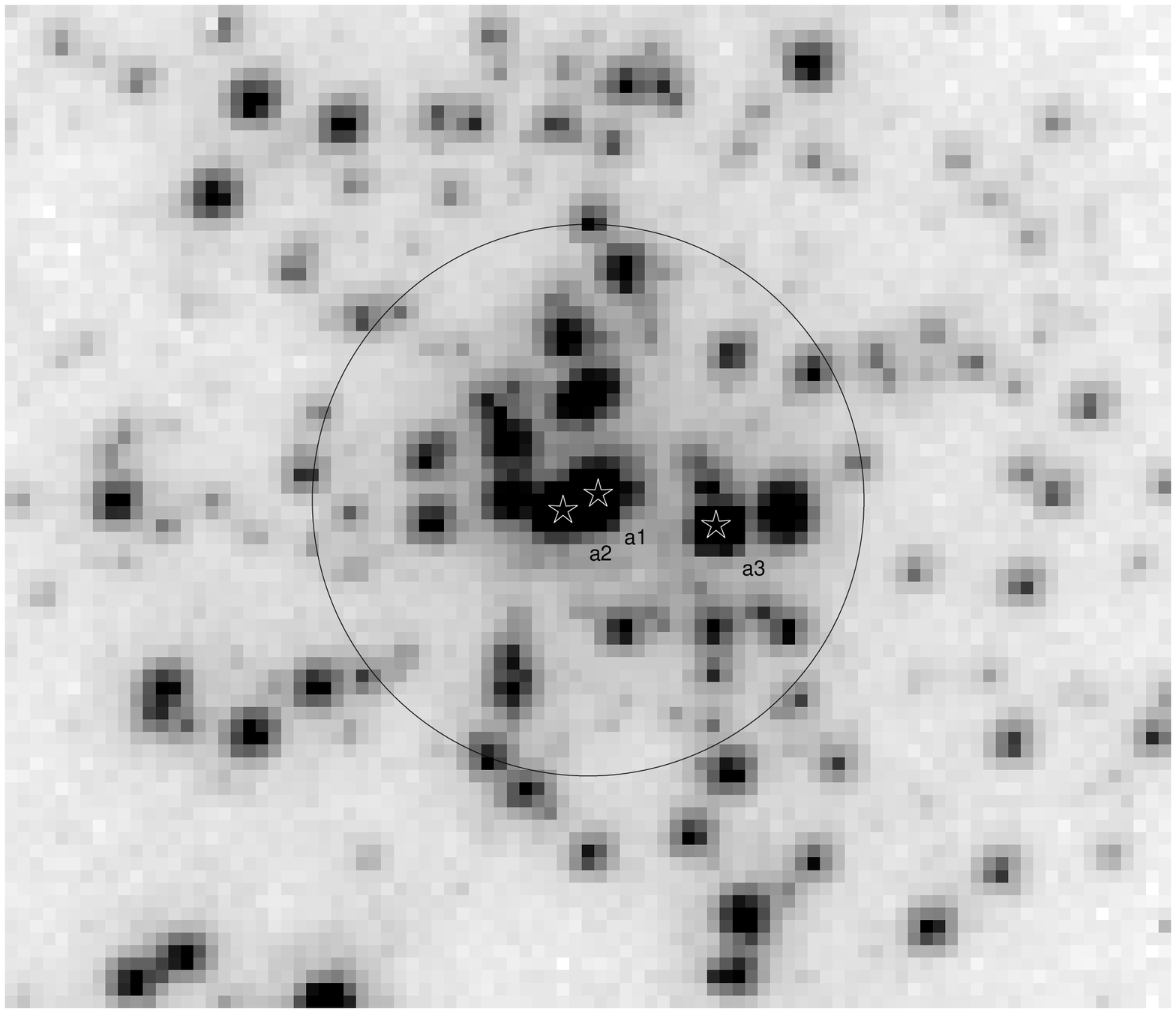}
\caption[top]{PC (WFPC2) F555W image of R136. The three most massive (WNha) objects are marked and a circle with radius of 1\arcsec\ centered on 
the cluster has been drawn.  A square-root scale between 0 and 400 counts has been used in order to show both the bright and the dim stars. 
The pixel size is 0\farcs0455, the field size is 4\farcs23$\times$3\farcs64 (93 px $\times$ 80 px), and the vertical direction is 
37.8\degr\ West of North. The physical size (1.0 pc $\times$ 0.9 pc) is similar to that of Fig.~\ref{ngc3603_f550m}.}
\label{r136_f555w}
\end{figure}	

	I have applied the results of the experiments above to two MYCs in the Local Group, NGC 3603 
(Fig.~\ref{ngc3603_f550m}) and R136 (Fig.~\ref{r136_f555w}), using HST data (HRC/ACS in the first case, PC/WFPC2 in the second one). 

	For NGC 3603 chance superpositions are a very small effects: most point sources in the HRC image are indeed single stars or 
unresolved real multiples. This implies that the massive-star AMF slope (between $-1.6$ and $-2.0$) is indeed very close to the IMF slope.
The NGC 3603 observations indicate that three stars have masses above 120 M$_\odot$, which is precisely the number of apparently 
ultramassive stars predicted in the first experiment if the real stellar upper mass limit is 120 M$_\odot$. Two of those three 
objects are indeed spectroscopic binaries (\citealt{Moffetal04}; Schnurr \& Moffat, private communication) and there is a relatively
separation gap where the third one could have an undetected (by spectroscopic or visual means) companion. 

	For the WFPC2 HST data chance superpositions play a more important role: near the core each apparent point source is expected to be composed
of the superposition of $\sim 2^7$ individual stars or $\sim 2^6$ binaries\footnote{Of course, most of the objects along a given line of
sight will be low-mass stars}. Such a large number makes the measurement of an IMF slope at the core impossible with the current data (the 
situation could change in the near future with new HST data, though). R136 also contains three stars with an observed mass of 120 M$_\odot$ 
or above. However, the existence of those three objects is also compatible with them being composed of $\sim 100$ stars with masse below 
120 M$_\odot$.

	Therefore, the conclusion from the analysis of the NGC 3603 and R136 data is that the case for a stellar upper mass limit is 
reinforced and that a value as low as 120 M$_\odot$ for solar metallicity is still possible.

\section{Random uncertainties and the gamma bias}

	A third effect that we plan to analyze in the future is that of random uncertainties in the determination of the IMF.
The massive-star IMF should be a monotonically decreasing function so one would expect that, given an observed mass, the
true mass would be more likely to be lower than higher than that value because in the parent
distribution there are more stars with lower than with higher masses. Such a ``diffusion'' from lower to higher masses should create an AMF
flatter than the real IMF, thus creating a ``gamma bias'' towards higher (lower in absolute value) values of the slope, an effect similar to the
one created by the Malmquist bias for luminosities or by the Lutz-Kelker bias for distances.

\bibliographystyle{spr-mp-nameyear-cnd}  
\bibliography{general.bib}               

%

\end{document}